\documentclass[10pt,aps,prd,amsmath,amssymb,superscriptaddress,twocolumn,nofootinbib,showpacs,preprintnumbers,amsfonts,notitlepage]{revtex4-1}

\usepackage[export]{adjustbox}
\usepackage[english]{babel}
\usepackage[utf8]{inputenc}
\usepackage{color,graphicx}
\usepackage{bm,bbm}
\usepackage{enumerate}
\usepackage{hyperref}
\usepackage{mathtools}
\usepackage{mathrsfs}
\usepackage{mciteplus}
\usepackage{slashed}
\usepackage{soul}
\usepackage{wrapfig}
\usepackage{xspace}
\usepackage[shortlabels]{enumitem}
\usepackage{orcidlink}
\usepackage[capitalise]{cleveref}
\usepackage{txfonts}  
\usepackage{relsize}

\newcommand{\defeq}{\overset{\mathrm{def}}{=\joinrel=}}

\renewcommand{\r}{\bm{r}}
\newcommand{\x}{\bm{x}}
\newcommand{\y}{\bm{y}}
\newcommand{\z}{\bm{z}}

\newcommand{\one}{\mathbbm{1}}

\newcommand{\Jc}{\mathcal{J}}

\newcommand{\Nc}{\mathcal{N}}
\newcommand{\Oc}{\mathcal{O}}

\DeclareMathOperator{\tr}{\text{Tr}}

\renewcommand{\min}{{\text{min}}}
\newcommand{\bra}{ \langle }
\newcommand{\ket}{ \rangle }

\newcommand{\der}{\partial}
\newcommand{\bcol}{\left[ \begin{array}{c}}
\newcommand{\ecol}{\end{array} \right]}
\newcommand{\beq}{\begin{eqnarray}}
\newcommand{\eeq}{\end{eqnarray}}

\newcommand{\ie}{{i.e.}\xspace}

\newcommand{\kev}{\ensuremath{{\mathrm{\,ke\kern -0.1em V}}}\xspace}
\newcommand{\mev}{\ensuremath{{\mathrm{\,Me\kern -0.1em V}}}\xspace}
\newcommand{\gev}{\ensuremath{{\mathrm{\,Ge\kern -0.1em V}}}\xspace}
\newcommand{\tev}{\ensuremath{{\mathrm{\,Te\kern -0.1em V}}}\xspace}

\usepackage{mfirstuc} 
\newcommand{\addReviewer}[2]{
  \expandafter\newcommand\csname #1\endcsname[1]{{\bf \color{#2} \capitalisewords{#1}:\,##1}}
  \expandafter\newcommand\csname #1cor\endcsname[2]{{\color{#2} \capitalisewords{#1}:\,\st{##1}{\bf ##2}}}
  \expandafter\newcommand\csname #1color\endcsname{#2}
}

\usepackage{soul,xcolor}

\definecolor{royalblue}{rgb}{0.255, 0.412, 0.882}

\definecolor{chromeyellow}{rgb}{1.0, 0.65, 0.0}
\definecolor{DodgeBlue}{rgb}{0.118, 0.565,1.000}
\definecolor{asparagus}{rgb}{0.53, 0.66, 0.42}
\definecolor{cardinal}{rgb}{0.77, 0.12, 0.23}
\definecolor{cadmiumgreen}{rgb}{0.0, 0.42, 0.24}
\definecolor{applegreen}{rgb}{0.55, 0.71, 0.0}
\definecolor{amethyst}{rgb}{0.6, 0.4, 0.8}

\addReviewer{wrong}{red}
\addReviewer{good}{green}

\addReviewer{sebastian}{cardinal}
\addReviewer{wyatt}{amethyst}
\addReviewer{arkaitz}{asparagus}
\addReviewer{eric}{cadmiumgreen}
\addReviewer{robert}{royalblue}
\addReviewer{cesar}{magenta}
\addReviewer{adam}{red}

\hypersetup{ 
    pdfnewwindow=true,
    colorlinks=true,
    linkcolor=royalblue,
    citecolor=royalblue,
    filecolor=royalblue,
    urlcolor=royalblue 
} 

\newcommand{\uw}{Department of Physics, University of Washington, WA 98195, USA}
\newcommand{\barc}{Departament de F\'isica Qu\'antica i Astrof\'isica and Institut de Ci\'encies del Cosmos, Universitat de Barcelona, E-08028 Barcelona, Spain}
\newcommand{\ceem}{Center for  Exploration  of  Energy  and  Matter,  Indiana  University,  Bloomington,  IN  47403,  USA}
\newcommand{\indiana}{Physics  Department,  Indiana  University,  
Bloomington,  IN  47405,  USA}
\newcommand{\jlab}{Theory Center, Thomas  Jefferson  National  Accelerator  Facility,  Newport  News,  VA  23606,  USA}
\newcommand{\odu}{Department of Physics, Old Dominion University, Norfolk, VA 23529, USA}
\newcommand{\uned}{Departamento de F\'isica Interdisciplinar, Universidad Nacional de Educaci\'on a Distancia (UNED), E-28040 Madrid, Spain}
\newcommand{\pitt}{Department of Physics and Astronomy, University of Pittsburgh, Pittsburgh, PA 15260, USA}
\newcommand{\gw}{Department of Physics, The George Washington University, Washington, DC 20052, USA}
\newcommand{\ucb}{Department of Physics, University of California, Berkeley, CA 94720, USA}
\newcommand{\lbnl}{Nuclear Science Division, Lawrence Berkeley National Laboratory, Berkeley, CA 94720, USA}

\begin{document}


\preprint{JLAB-THY-24-4145}

\title{Coulomb confinement in the Hamiltonian limit}

\author{Sebastian~M.~Dawid\orcidlink{0000-0001-8498-5254}}
\email[email: ]{dawids@uw.edu}
\affiliation{\uw}
\author{Wyatt~A.~Smith\orcidlink{0009-0001-3244-6889}}
\email[email: ]{wyatt.smith@gwu.edu}
\affiliation{\gw}
\affiliation{\ucb}
\affiliation{\lbnl}
\author{Arkaitz~Rodas\orcidlink{0000-0003-2702-5286}}
\affiliation{\odu}
\affiliation{\jlab}
\author{Robert~J.~Perry\orcidlink{0000-0002-2954-5050}}
\affiliation{\barc}
\author{C\'esar~Fern\'andez-Ram\'irez\orcidlink{0000-0001-8979-5660}}
\affiliation{\uned}
\author{Eric~S.~Swanson\orcidlink{0000-0001-7394-9215}}
\affiliation{\pitt}
\author{Adam~P.~Szczepaniak\orcidlink{0000-0002-4156-5492}}
\affiliation{\jlab}
\affiliation{\indiana}
\affiliation{\ceem}


\begin{abstract}
The Gribov--Zwanziger scenario attributes the phenomenon of confinement to the instantaneous interaction term in the QCD Hamiltonian in the Coulomb gauge. For a static quark-antiquark pair, it leads to a potential energy that increases linearly with the distance between them. Lattice studies of the SU(2) Yang--Mills theory determined the corresponding (Coulomb) string tension for sources in the fundamental representation, $\sigma_{C}$, to be about three times larger than the Wilson loop string tension, $\sigma_F$. It is far above the Zwanziger variational bound, $\sigma_C \geq \sigma_F$. We argue that the value established in the literature is artificially inflated. We examine the lattice definition of the instantaneous potential, find the source of the string tension's enhancement, and perform its improved determination in SU(2) lattice gauge theory. We report our conservative estimate for the value of the Coulomb string tension as $\sigma_C/\sigma_F = 2.0 \pm 0.4$ and discuss its phenomenological implications.
\end{abstract}

\date{\today}
\maketitle


\section{Introduction}
\label{sec:intro}

The Coulomb gauge Hamiltonian formalism has long been used to study non-perturbative aspects of quantum chromodynamics (QCD), including color confinement. According to the \mbox{Gribov--Zwanziger} scenario~\citep{Gribov:1977wm, Zwanziger:1991gz, DellAntonio:1991mms, Zwanziger:1998}, quark confinement\footnote{We define confinement as an area law of the Wilson loop, producing a linearly rising potential between static charges~\citep{Wilson:1974}.} manifests explicitly in the instantaneous, chromo-electric part of the Coulomb-gauge SU($N_c$) \mbox{Yang--Mills} (YM) Hamiltonian. One may interpret this interaction as a non-Abelian generalization of the familiar Coulomb potential between electric charges separated by a distance $r$. However, unlike in quantum electrodynamics (QED), the non-Abelian instantaneous part of the QCD Hamiltonian is a complicated function of the background gauge fields. As a result, at large separations between color sources, it is expected that the associated potential rises linearly ~\citep{Cucchieri:2000, Szczepaniak:2001, Szczepaniak:2003ve, Zwanziger:2003i, Greensite:2004ur, Campagnari:2009wj, Leder:2011, Golterman:2012, Reinhardt:2017, Cooper:2018}. This potential, which we refer to as $V_{\text{C}}(r)$, is defined as interaction energy between a static quark-antiquark pair annihilated immediately after its creation, i.e. before dynamical quark-gluon interactions anti-polarize the QCD vacuum around the pair. A state of energy equal to $V_C$ is referred to as ``bare'' or ``stringless,'' meaning it does not describe a flux tube-like gauge field distribution connecting the quarks.

In general, with the static color charges in representation $D$ of SU($N_c$) and assuming large $r$, it was shown in~\citep{Zwanziger:2003i} that the relation,
\begin{align}
\label{eq:tension_bound}
V_D(r) \leq V_{\rm C}(r) \, ,
\end{align}
holds, where $V_D(r) \approx \sigma_D \, r$ is the ground-state potential of a dressed quark-antiquark pair as computed from the expectation value of a large Wilson loop.\footnote{As originally stated by Zwanziger, the right-hand side of \cref{eq:tension_bound} contains the quadratic Casimir invariant $C_D$. We have absorbed this quantity into our definition of $V_C$ to match the conventions of recent literature on the subject.} Here, $\sigma_D$ is the ground-state string tension. We refer to the corresponding linear term in $V_C(r) \approx \sigma_C \, r$ as the non-Abelian Coulomb string tension $\sigma_C$. From \cref{eq:tension_bound}, it follows that if the physical potential confines, so does the non-Abelian Coulomb one. Moreover, if the string tensions were the same, one could argue that the bare and the ground state would be identical.

Various computations of the non-Abelian Coulomb potential in the fundamental representation of the SU(2) and SU(3) theories have found that the Coulomb string tension is about three to four times larger than the Wilson string tension, \ie, \mbox{$\sigma_C/ \sigma_F\gtrapprox 3$}~\citep{Greensite:2003, Greensite:2004ke, Nakamura:2005ux, Nakagawa:2006fk, Iritani:2010mu, Greensite:2014bua, Burgio:2015hsa, Greensite:2015nea}. Thus, it is far above Zwanziger's lower bound, indicating a significant difference between the bare and the ground state. The considerable disparity between the string tensions poses a problem for the viability of phenomenological meson models using the Coulomb-gauge Hamiltonian approach~\citep{Guo:2008yz, Guo:2014zva, Amor-Quiroz:2017jhs, Farina:2020slb, Abreu:2020ttf, Swanson:2023zlm, Farina:2023oqk}; it necessitates the inclusion of many additional string-like gauge fields in constructing the ground-state. Possible explanations exist for the dynamical mechanism that reduces the enhanced instantaneous interaction energy to the ground-state from the bare one ~\citep{Greensite:2001nx, Greensite:2009mi, Ostrander:2012kz, Greensite:2014bua}, but it remains unknown what causes such a discrepancy between the two potentials. Recent lattice simulations of the chromoelectric energy density distribution in the ``stringless'' Coulomb state were inconclusive~\citep{Chung:2017, Dawid:2019vhl}, indicating a need for further study.

Lattice computations of the non-Abelian Coulomb potential are notoriously challenging. Two discretized definitions of this quantity were proposed: one defined as a correlator between short time-like links~\citep{Greensite:2003}, while the other based on resolving the Gauss constraint and expressing the instantaneous part of the temporal gluon propagator in terms of the \mbox{Faddeev--Popov} operator~\citep{Cucchieri:2000gu, Quandt:2008zj, Burgio:2012bk}. Although equivalent in the continuum Hamiltonian formulation, they differ significantly on a finite lattice. In addition to exhibiting different discretization and finite-volume effects, they differ because the observables of interest are defined in the Coulomb gauge and may be affected differently by Gribov copies. For example, a lower value for the Coulomb string tension was found~\citep{Voigt:2008rr, Nakagawa:2010eh, Burgio:2012bk} using the latter definition; however, as realized in Refs.~\citep{Heinzl:2007cp, Burgio:2016nad}, that string tension depends strongly on the gauge-fixing procedure. According to~\citep{Burgio:2016nad}, the former definition—studied in this article—is most likely unaffected by this difficulty, although more studies are necessary.

Nevertheless, the lattice version of the non-Abelian Coulomb potential, as defined in Ref.~\citep{Greensite:2003}, suffers from other problems, both formal and computational. As discussed in the following section, to obtain the Coulomb string tension, one must study the logarithmic derivative of a correlator defined as the expectation value of an operator of vanishing temporal extent, $t \to 0$. It requires analyzing the short-distance behavior of a correlation function for which large lattice artifacts are to be expected. Thus, one must carefully investigate the dependence of the extracted Coulomb string tension on the temporal lattice spacing, $a_t$. Crucially, the limit $t \to 0$ is achievable only in the Hamiltonian limit, \ie, when $a_t \to 0$. In practice, measuring the Coulomb potential for several ensembles with sufficiently fine spacings while accounting for all relevant systematic effects remains an open issue. Attempts to explore this limit often delivered inconclusive results, showing, for example, a significant dependence of the ratio $\sigma_C/\sigma_F$ on the inverse coupling $\beta$~\citep{Greensite:2003}.

Additional theoretical challenges might have influenced these computations. In particular, we assert that the commonly assumed form of this derivative may be inappropriate for application to small Euclidean times, where the standard power-counting arguments implicit in Symanzik's effective action approach do not hold. At a small temporal extent, $t \approx a_t$, relevant to our observable of interest, it is crucial to account for non-negligible cutoff effects. We delay the discussion of this technical point until further in the article. The most important consequence of this problem is a potential artificial enhancement of the Coulomb string tension, which can exceed the correct value by more than 100\%~\citep{Smith:2023pua}.

In this work, we present straightforward solutions to these issues and perform improved SU(2) lattice simulations on several anisotropic lattices at different couplings. We estimate the value of the Coulomb string tension and present a conservative estimate for the ratio \mbox{$\sigma_C/\sigma_F = 2.0 \pm 0.4$}, closer to Zwanziger's bound than was previously determined. This improvement offers hope of resolving several puzzles in the \mbox{Gribov--Zwanziger} confinement scenario and of making a connection with the Casimir scaling phenomenon, observed in lattice simulations~\citep{Ambjorn:1984mb, Greensite:1979yn, Olesen:1981zp, Deldar:1999vi, Bali:2000un, Piccioni:2005un, Cardoso:2011cs}.

This article is organized as follows. In~\cref{sec:potential}, we define the non-Abelian Coulomb potential and discuss its lattice equivalent. We highlight problems with the commonly used version of this observable and offer an approach that resolves them. In~\cref{sec:results}, we describe the details of our lattice setup and statistical analysis. We present the results of the numerical computation and the Hamiltonian limit of the non-Abelian Coulomb potential. In~\cref{sec:con}, we summarize our findings and discuss potential improvements of this work. We include four technical Appendices. In~\cref{app:enhancement}, we give a simple argument explaining the mechanism responsible for enhancing previous measurements of the non-Abelian Coulomb potential. In~\cref{app:anisotropy}, we discuss the details of anisotropic lattices used for the computation. In~\cref{app:MA}, we discuss the statistical analysis performed on the lattice data. Finally, in~\cref{app:Wilson_limit}, we discuss our computation of the Wilson loop string tension.

\section{ Non-Abelian Coulomb potential }
\label{sec:potential}

We begin by describing the origin of the non-Abelian Coulomb potential in the Hamiltonian of the SU($N_c$) YM theory with static quarks. The static quark-antiquark creation operators are represented at space-time point \mbox{$(t=0,\bm{r})$} by $q_a(\r)$ and \mbox{$ \bar{q}_a(\r)$} respectively. The bare quark-antiquark state is defined as \mbox{$|\Psi_{q\bar{q}} \ket = \bar{q}_a(\bm{0}) \, q_a(\r) \, |\Omega \ket$} where the YM vacuum is denoted by $|\Omega\ket$. Formally, one may split the Hamiltonian in the Coulomb gauge into four parts, \mbox{$H_{\rm YM} = H_{q}+H_{g}+ H_{qg} + H_C$}, where the first three terms contain contributions from free quarks, free gluons, gluon-gluon interactions, and quark-gluon interactions respectively. The final term, $H_C$, describes the instantaneous interaction between color charges and is given by~\citep{Schwinger:1962wd, Christ:1980ku, Reinhardt:2017},
\begin{align}
H_C = \frac{1}{2} \! \int \! d\x \, d\y \, 
\Jc^{-1/2} \, \rho^a(\x) \, \Jc^{-1/2} \,
K^{ab}_{\x \y} \, \Jc^{-1/2}
\, \rho^b(\x) \, \Jc^{-1/2} \,  ,
\end{align}
where $\rho^a(\x) = g q^\dag(\x) t^a q(\x)$ is the color charge density operator of quarks, \mbox{$\Jc=\text{det}(M)$} is the determinant of the \mbox{Faddeev--Popov} operator present due to quantization in curvilinear coordinates on the gauge manifold, \mbox{$M^{ab}_{xy} = -\der^i \, D_i^{ab} \, \delta^{(4)}(x-y)$}, and the covariant derivative is \mbox{$D_\mu^{ab} = \delta^{ab} \,\der_\mu - g \, f^{abc} \, A^c_\mu$}. Implicit summation over repeated discrete indices is assumed. 

Similarly to QED, the interaction kernel is obtained by resolving the Gauss-law constraint. The result is,
\begin{align}
K^{ab}_{\x\y} = [M^{-1} \, (-\nabla^2) \, M^{-1} ]^{ab}_{\x\y} \, .
\end{align} 
In the Abelian limit, $M\to-\nabla^2$, and the kernel reproduces the familiar $1/r$ potential. The instantaneous chromo-electric interaction in a non-Abelian theory is affected by possible gluon exchanges and becomes non-perturbatively enhanced.\footnote{As argued by Gribov, Zwanziger, and others, the eigenvalues of $M^{-1}$ become large near the surface of the so-called Fundamental Modular Region~\citep{Gribov:1977wm, Greensite:2011zz} which contains the allowed field configurations in the gauge-fixed theory. Due to the high dimensionality of this space, most of its volume resides near the surface, and consequently, the non-Abelian Coulomb potential is enhanced at large separations, leading to confinement.} The Coulomb potential is defined (up to an infinite self-interaction constant) as a matrix element, 
\begin{align}
\label{eq:coulomb_pot_hamiltonian}
V_C(r) = \bra \Psi_{q\bar{q}}| \, H_C \, | \Psi_{q\bar{q}} \ket = \bra \Psi_{q\bar{q}}| \, H \, | \Psi_{q\bar{q}} \ket \, .
\end{align} 
The equality between the expectation value of the full Hamiltonian and that of the Coulomb term results from the instantaneous nature of the interaction. At a given instant, for a finite separation between quarks, terms $H_q$, $H_g$, and $H_{qg}$ all have vanishing matrix elements and do not contribute to the interaction energy. 

Based on this fact, in Ref.~\citep{Greensite:2003}, the authors proposed to use the expectation value of the Euclidean time evolution operator on the $|\Psi_{q\bar{q}}\ket$ state,
\begin{align}
V_C(r) = - \lim_{t \to 0} \frac{d}{dt} \log \bra \Psi_{q\bar{q}}| \, \text{e}^{-t H} \, | \Psi_{q\bar{q}} \ket \, ,
\end{align}
for computing the potential in the lattice Monte Carlo simulation. For this purpose, they introduced the correlation function,
\begin{align}
\label{eq:cont-correlator}
G(r,t) = \frac{1}{2} \, \langle \text{Tr} \big[ L^\dag(\bm{0}, t) \, L(\r, t) \big] \rangle \, .
\end{align}
where $\langle \cdot \rangle$ represents the vacuum expectation value, and $L(\x,t)$ is a temporal Wilson line of length $t$,
\begin{align}
\label{eq:cont-wil-line}
L(\bm{r}, t) = \text{P} \exp \left(\int_0^t dt' \, A_4(\bm{r}, t') \right) \, . 
\end{align}
In the static limit, $L(\bm{r}, t)$ describes the creation, Euclidean-time propagation, and annihilation of a single quark. From the definition of the Wilson line, note that at $t=0$ this correlation function is normalized to unity,
\begin{equation}
\label{eq:normalization}
G(r,0)=\frac{1}{2} \, \langle \text{Tr} \big[ L^\dag(\bm{0}, 0) \, L(\r, 0) \big] \rangle = 1\,.
\end{equation}
One obtains the Coulomb potential from
\begin{align}
\label{eq:cont-coul-pot}
V_C(r) = \lim_{t \to 0} V(r,t) \defeq \lim_{t \to 0} \bigg( - \frac{d}{dt} \log[ G(r,t) ] \bigg) \, .
\end{align}
    
The limit $t \to 0$ might seem unusual when viewed from the perspective of the typical $t \to \infty$ regime studied in hadronic spectroscopy on the lattice. The latter allows one to access the ground-state energy of interest by using the large-time dominance of the lowest-energy state's exponential in the spectral decomposition of $G(r,t)$. In fact, the quantity $\lim_{t \to \infty} V(r,t)$ reproduces the well-known ground-state quark-antiquark potential extracted from the asymptotic behavior of the large Wilson loop~\citep{Greensite:2003}. One can intuitively understand this correlator as describing a process of equilibration of the bare state with time. At $t = 0$, a bare quark-antiquark state is produced with its energy initially given only by the contribution of the instantaneous part of the Coulomb gauge Hamiltonian. As time evolves, interactions of the bare state with the virtual gluons in the YM vacuum allow it to ``thermalize'' to the ground state of the theory. Hence, the quantity $V(r,t)$ interpolates between the non-Abelian Coulomb potential and the physical (ground-state) one, and the corresponding time-dependent string tension $\sigma(t)$ satisfies, $\sigma_C \geq \sigma(t) \geq \sigma_F$.

We now consider the gauge-fixed theory in a hypercubic lattice of volume $L^4$, where \mbox{$L = N a_s = (\xi N) a_t$} and the integer $N$ determines lattice size in units of the spatial and temporal spacings, $a_s$ and $a_t$. Here, \mbox{$\xi = a_s/a_t$} is the renormalized anisotropy, and we assume periodic boundary conditions in all directions. The Wilson lines then take on the form, 
\begin{align}
\label{eq:latt-wilson-line}
L_{\rm lat}(t,\r) = U_4(\r,0)\, U_4(\r,a_t)\, \dots\, U_4(\r,a_t (T-1)) \, ,
\end{align}
where $t = a_t T$, $r = a_s R$, with $T$ and $R$ integers, and $U_\mu(x)$ is a gauge link variable at position \mbox{$x=(\bm{x},x_4)$} pointing in the $\mu \in [1,4]$ direction. One represents the correlation function with the lattice approximant,
\begin{align}
\label{eq:latt-corr}
G_{\rm lat}(r,t; a_t, a_s, N) = \frac{1}{2} \, \big\langle \text{Tr} \big[ L_{\rm lat}^\dag(\bm{0}, t) \, L_{\rm lat}( \r, t) \big] \big\rangle \, ,
\end{align}
where we explicitly indicate dependence on the lattice spacings and size. The most direct approach to obtaining the Coulomb potential is to replace the time derivative in~\cref{eq:cont-coul-pot} with a finite difference. By doing so, we find the time-dependent potential,
\begin{align}
\label{eq:latt-potential-time}
V_{\rm lat}(r,t; a_t, a_s, N) = \frac{1}{a_t} \log \frac{G_{\rm lat}(r,t; a_t, a_s, N)}{G_{\rm lat}(r,t+a_t; a_t, a_s, N)} \ .
\end{align}
Reference~\citep{Greensite:2003} proposes to take the $t\to 0$ limit by fixing $t = 0$ and assuming $G_{\rm lat}(r,0; a_t, a_s, N) = 1$, as suggested by Eq.~\eqref{eq:normalization}. It gives the lattice equivalent of the continuum Coulomb potential,
\begin{align}
\label{eq:latt-coul-wrong}
\widetilde{V}_{\rm lat}(r; a_t, a_s, N) &= - \frac{1}{a_t} \log \Big\langle \frac{1}{2} \text{Tr} [ U_4^\dag( \bm{0},0) U_4( \bm{r},0) ] \Big\rangle \, ,
\end{align}
as implemented in many lattice studies.

However, it is crucial to note that data obtained from lattice QCD (LQCD) studies cannot be directly used to verify the assumption of unit normalization of the lattice approximant of the correlator at finite spacing, since the correlator defined by Eq.~\eqref{eq:latt-corr} can only be computed for temporal separations $T\geq 1$. 

In addition, observables computed using LQCD generically exhibit systematic effects due to discretization. Such artifacts are ubiquitous and are removed by studying the continuum limit of LQCD. However, at finite lattice spacing, from dimensional arguments, one anticipates the presence of a class of lattice artifacts that modify the time-dependence of the correlation function measured on the lattice~\citep{DellaMorte:2008xb, Harris:2021azd, Ce:2021xgd}. We write, for a generic correlator on an isotropic lattice in an infinite volume,
\begin{align}
\label{eq:discrete-corrections-1}
G_{\rm lat}(r,t;a) = G_0(r, t;a) + \sum_{n=1} \left(\frac{a}{t}\right)^n G_n(r, t;a) \, ,
\end{align}
where $G_n(r, t;a)$ are regular and finite at $t=0$. After multiplicative renormalization, the continuum limit of this correlator may be obtained by studying $\lim_{a\to0}G_{\rm lat}(r,t; a)=\lim_{a\to0}G_0(r,t; a)$. Thus, the $G_0$ term determines the Coulomb potential.

The terms $\left(a/t\right)^n G_n(r,t; a)$ for $n\geq1$ parameterize lattice artifacts which are enhanced for small-$t$ due to the explicit factors of $\left(a/t\right)^n$, and modify the normalization of the lattice correlation function at $t=0$. Consequently, the assumption of unit normalization at $t=0$ of the correlation function at finite lattice spacing is not justified\footnote{Stated differently, the lattice approximant of $G_{\rm lat}(r,t; a)$ has a normalization that is time- and spacing-dependent, $\Nc(t, a) \neq 1$. It approaches unity, $\Nc(t, a) \to 1$, when one performs the ordered limit $a \to 0$ and then $t \to 0$ in \cref{eq:latt-corr}. Since the definition of the Coulomb potential in Eq.~\eqref{eq:latt-coul-wrong} violates this order, it suffers from non-vanishing terms $(a/t)^n$ contributing to the normalization at any $a$.} and may cause an enhancement in the measured potential as $t$ decreases~\citep{Smith:2023pua}. We explain this point further in~\cref{app:enhancement}.

In principle, this issue may be removed by first extrapolating the correlation function point-by-point to the continuum and then analyzing the small-time dependence of the resulting renormalized continuum correlation function. This approach is beyond the scope of this work. Instead, we note that the assumption of unit normalization of the correlation function at $t=0$ is not required to extract the Coulomb potential. We therefore analyze the time dependence of Eq.~\eqref{eq:latt-corr} and directly fit the correlator near $t=0$ with a proper continuous model, $G_{\rm mod}(r,t; \vec{\alpha})$, where $\vec{\alpha}$ is a vector of free parameters that depend on $a_t, a_s,$ and $N$. The complete model includes terms corresponding to the lattice artifacts in~\cref{eq:discrete-corrections-1}, potentially with a divergence at $t=0$. The model takes the form,
\begin{align}
\label{eq:discrete-corrections-2}
G_{\rm mod}(r,t;\vec{\alpha}) = G_{0,\rm mod}(r, t;\vec{\alpha}) + \sum_{n=1} \left(\frac{a}{t}\right)^n G_{n,\rm mod}(r, t;\vec{\alpha}) \, .
\end{align}
Once the model parameters are constrained, we can identify $G_{0,\rm mod}(r, t;\vec{\alpha})$ as the term relevant to the extraction of the Coulomb potential. It is important to note that this term still contains finite-volume and finite-spacing dependencies, which may be removed by studying the continuum limit. Since the leading lattice artifacts arise from the temporal discretization, we assume it is sufficient to study the limit where the temporal spacing becomes zero, $a_t \to 0$. It is approached by studying anisotropic lattices with increasing anisotropy, as $\xi \to \infty$ corresponds to the desired $a_t\to 0$ limit. Thus, we study,
\begin{align}
\label{eq:ham-limit}
V_C(r; a_s, N) = \lim_{\substack{\xi \to \infty \\ L = {\rm const.}}}  \left( - \lim_{t \to 0} \frac{d}{dt} \log\big[ G_{0,\rm mod}(r,t; \vec{\alpha}) \big] \right) \, .
\end{align}

\section{Numerical results}
\label{sec:results}

In this section, we discuss our numerical implementation of the Monte Carlo simulation of the SU($N_c$) YM theory on the lattice and our results for the non-Abelian Coulomb potential using the improved approach to the observable.

\subsection{Lattice framework}
\label{sec:lattice}

We consider the Euclidean lattice formulation of the SU(2) YM theory with no dynamical fermions. The theory is defined via the Wilson action for anisotropic lattices,
\begin{align}
S = \frac{1}{2} \sum_x & \bigg[ \,
\beta_s \! \sum_{j>i=1}^3 \text{Re} \tr \Big(\one - U_{ij}(x) \Big) \nonumber \\
\label{eq:action}
& + \beta_t \sum_{i=1}^3 \text{Re} \tr \Big( \one - U_{i4}(x) \Big) \bigg] \, ,
\end{align}
where,
\begin{align}
U_{\mu\nu}(x) = U_\mu(x) \, U_{\nu}(x+\hat{\mu}) \, U^\dagger_\mu(x+\hat{\nu}) \, U^\dagger_\nu (x) \, ,
\end{align}
is a plaquette oriented in the $(\mu \nu)$ plane and placed at position $x = ( \bm{x},x_4)$. Here $\beta_s = \beta /\xi_0$ and $\beta_t = \beta \, \xi_0$ are coupling constants introduced to alter the physical size of the lattice in the spatial and temporal directions, respectively. The quantity $\xi_0$ is the bare anisotropy, tuned in the action to set the renormalized anisotropy to $\xi$. Details regarding the determination of the renormalized anisotropy and lattice spacings, $a_t \equiv a_t(\beta, \xi)$ and $a_s \equiv a_s(\beta, \xi)$, are given in~\cref{app:anisotropy}.

\begin{table}
\caption{Summary of the lattice ensembles used in this work. For each lattice size $N$, eight $\beta$s are computed in steps of 0.05. The fourth column provides the minimal number of distinct gauge configurations used for measurements.}
\label{tab:ensembles}
\begin{ruledtabular}
\begin{tabular}{ccccc}
$N$ & $\beta$ & $\xi$ &$\min \! \left(n_{\rm meas}\right)$ \\
\hline
$16$ & 2.25 to 2.60 & 1 to 5 & 1917 \\  
$24$ & 2.25 to 2.60 & 1 to 8 & 1557\\  
$32$ & 2.25 to 2.60 & 1 to 4 & 494
\end{tabular}
\end{ruledtabular}
\end{table}

In our simulations, we used lattices with dimensions $N^3 \times  \xi N$, where $\xi N$ is the number of lattice sites in the time direction. This choice approximately maintains the physical volume of each lattice for fixed $\beta$ as $\xi$ increases. We used ensembles with $N=16, 24, 32$ for couplings $\beta=2.25$ to $2.60$ and anisotropy $\xi$ up to $8$. We summarize the employed lattice ensembles in \cref{tab:ensembles}. The SU(2) lattices were iteratively updated using the heat-bath algorithm~\citep{Creutz:1980, Kennedy:1985}. For each $\beta$ and $\xi$, we used 100 independent lattices (Markov chains) considered equilibrated after about $n_{\rm therm} = 2000$ to $5000$ initial thermalization sweeps. After equilibration, we generated $5$ to $20$ more lattice configurations in every Markov chain, each separated by $200$ to $400$ thermalization sweeps. These were used for measurements, which resulted in about $n_{\rm meas} = 500$ to $2000$ different gauge configurations employed for this purpose.

Before each measurement, the field configuration was fixed to the Coulomb gauge by employing the local relaxation algorithm of Ref.~\citep{Schrock:2012fj} at each time slice. A configuration was assumed to be gauge-fixed when $\Delta F = |F_i-F_{i+1}| < 10^{-7}$, where
\begin{align}
F_i = \frac{1}{4 \xi N^2} \sum_{\mu=1}^3 \sum_x \text{Tr} \hspace{2pt} U_\mu(x) \, ,
\end{align}
is a value of the functional to be minimized after the $i$-th gauge fixing iteration. To increase the speed of the gauge fixing procedure, we implemented an over-relaxation method~\citep{Mandula:1990, Giusti:2001} with $\omega=1.75$. We did not find significant changes in the values of our observables when decreasing $\Delta F$ by several orders of magnitude below $10^{-7}$.

For each gauge-fixed configuration we measured the correlator defined in~\cref{eq:latt-corr}, $G_{\rm lat}(R a_s, T a_t; a_t, a_s, N)$, for every $ 1 \leq R,T \leq N/2$. For the ensembles with $N=16$, we measured up to $T=10$ to better constrain the $T$-dependence of the correlator. Each measurement involved an average over three spatial orientations, $\hat{\x}$, $\hat{\y}$, $\hat{\z}$, of the quark-antiquark separation vector, $\r$, and an average over all possible time and spatial translations of the $L_{\rm lat}^\dag(\bm{0},t) \, L_{\rm lat}(\bm{r},t)$ operator. 

\subsection{Fitting strategy}
\label{sec:fitting}

\begin{figure*}
\begin{tabular}{cc}
\includegraphics[width=\textwidth]{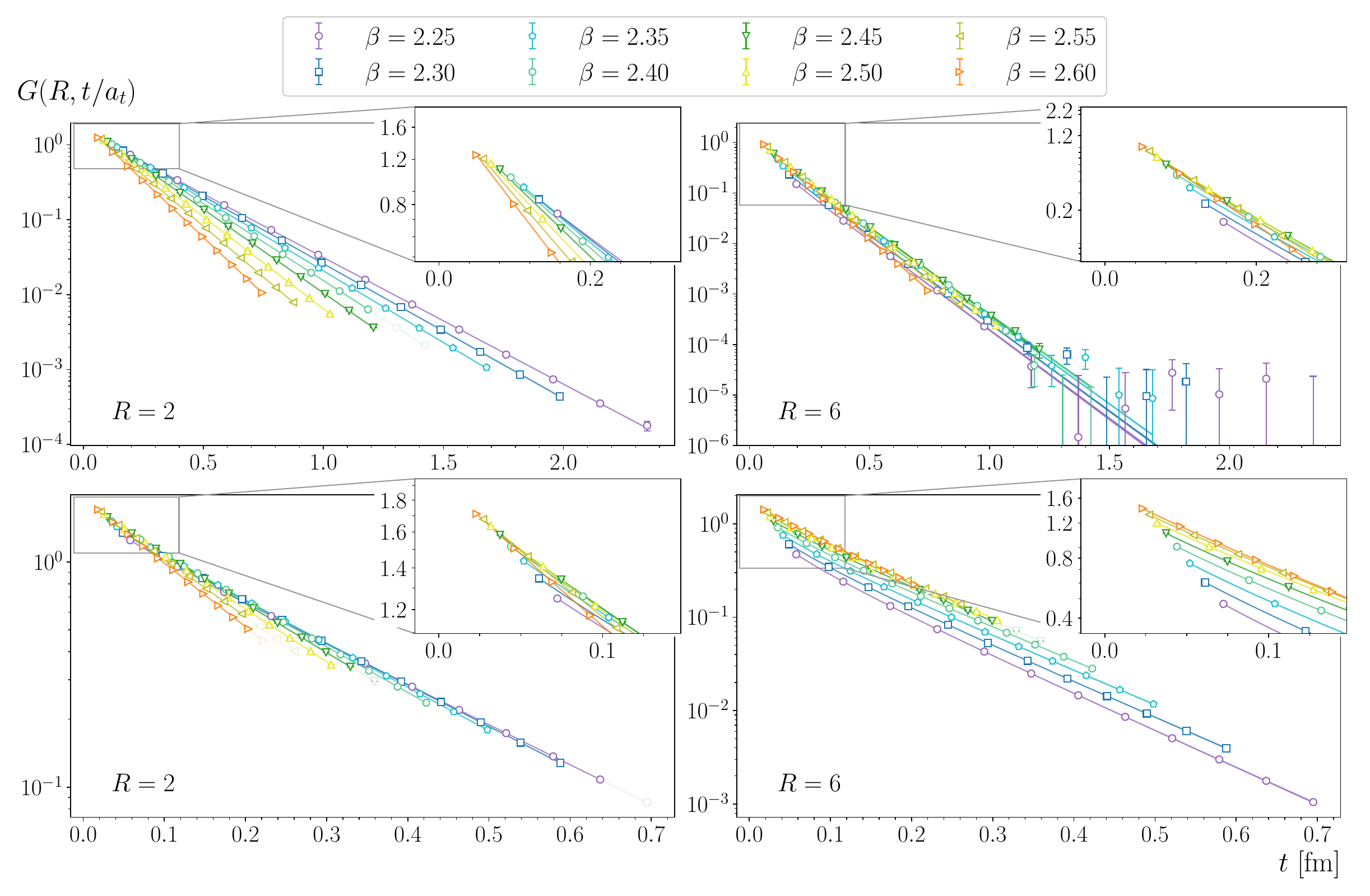} 
\end{tabular}
\caption{Time-dependence fits to our correlation functions for two values of $R$, for $N=24$ lattices with $\xi=1$ (top panels) and $\xi=4$ (bottom panels), for all $\beta$s used in this work. Shown is the best model selection fit to the data only (filled data points and error band). Transparency has been added to the region not fitted by the best model selection fit.}
\label{fig:C(t)_fits}
\end{figure*} 

\begin{figure*}
\includegraphics[width=0.9\textwidth]{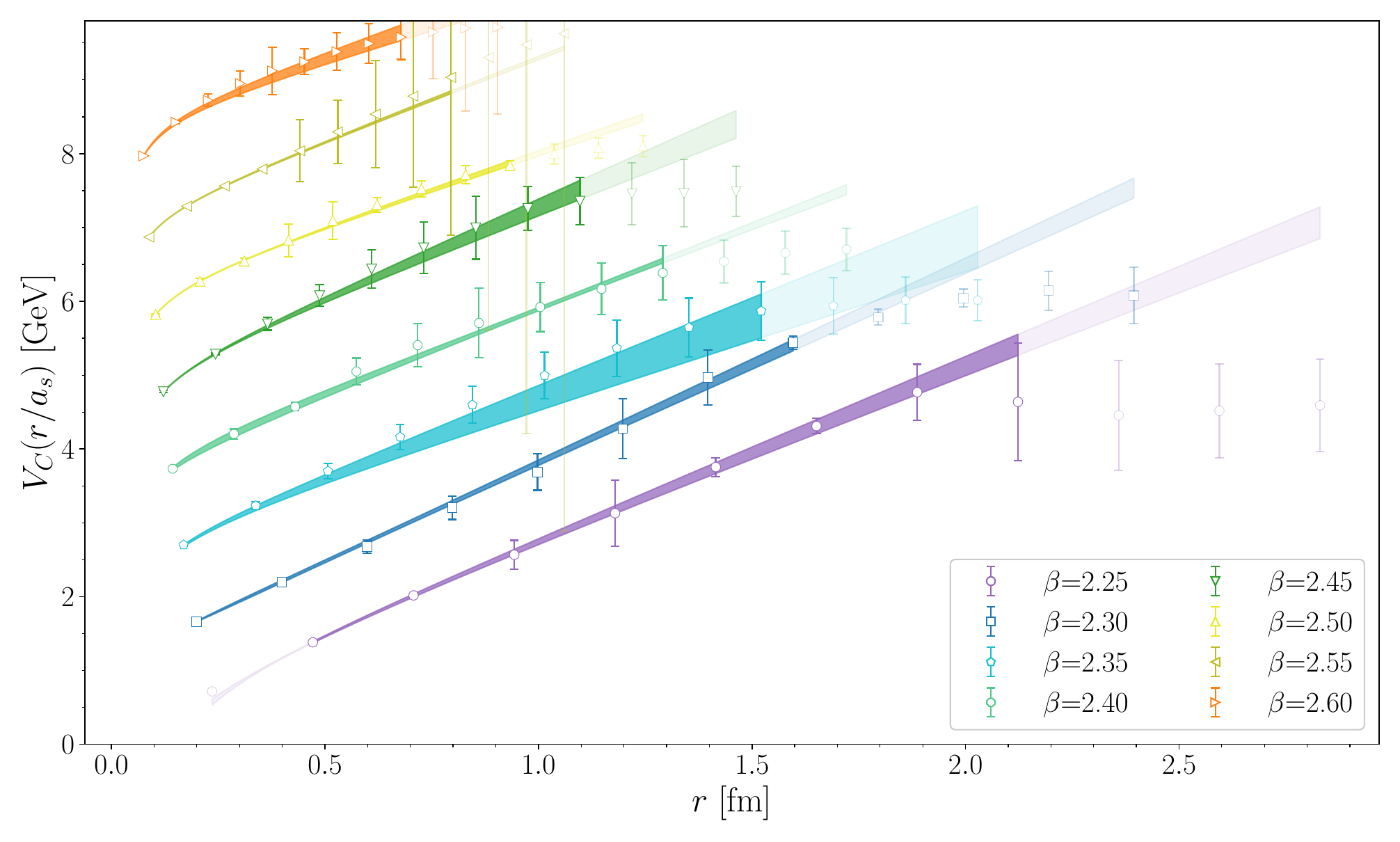}
\caption{Distance-dependent fits to our potential function for $N=24,\,\xi=8$. Shown for each value of $\beta$ is the best model fit to the data. The $y$-axis results for every $\beta=2.25+0.05\times n$ have been shifted by $n$ for clarity of presentation. Note that all the plots exhibit similar slopes, hinting at a string tension that is constant in $\beta$. Transparency has been added to the region not fitted by the best model fit.} \label{fig:V(r)_fits}
\end{figure*} 
 
We follow a multi-step procedure to extract the Coulomb string tension from the lattice approximant, \mbox{$G(R, T) \equiv G_{\rm lat}(R a_s, T a_t;a_t,a_s, N)$}. We first fit the small $t/a_t$ dependence of the correlation function to a range of different models at fixed $R$. We compute, within each of these models, the logarithmic derivative at $t=0$, as given in the parenthesis of~\cref{eq:ham-limit}. This allows us to construct the lattice approximation of the Coulomb potential, $V_C(R) \equiv V_{\rm lat}(R a_s;a_t,a_s, N)$. In the second step, we fit the $R$-dependence of the potential to obtain the Coulomb string tension, $\sigma_C(a_t,a_s,N)$. Finally, we fit the residual $a_t/a_s=1/\xi$ dependence to extrapolate the obtained quantity to $a_t \to 0$.

For the two first steps, the lattice data is considered correlated, and all our fits are performed in lattice units. The temporal evolution of our correlation function is fitted by minimizing the standard correlated $\chi^2$,
\begin{widetext}
\begin{align}
\chi^2=\sum_{ij}  
\Big( G(R,T_i)-G_{\rm mod}(R,T_i; \vec{\alpha}) \Big) \, \, \mathlarger{\Sigma_{ij}^{-1}} \, \, \Big( G(R,T_j)-G_{\rm mod}(R,T_j; \vec{\alpha}) \Big) \, ,
\label{eq:G_chi2}
\end{align}
\end{widetext}
where $\vec{\alpha}$ is the parameter vector, and $\Sigma_{ij}$ represents the covariance matrix between the samples $G(R,T_i)$ and $G(R,T_j)$. Fits to data sets with a truncated $T$-range were handled by truncating the covariance matrix before inverting it for insertion into \cref{eq:G_chi2}.\footnote{The same procedure is followed for fits to $V(R)$.} The errors and correlations of our observables are obtained via the jackknife procedure, as described in~\cref{app:MA}. Finally, model averaging~\citep{Jay:2020jkz} was employed to estimate the systematic uncertainty arising from model choice, and from the ranges of $R,\,T$ used for the fits presented in this work. 

\subsubsection{Fitting the $t$-dependence of the correlator}
\label{sec:time-dependence}

As discussed in~\cref{sec:potential} and~\cref{app:enhancement}, lattice artifacts of the form $(a_t/t)^n$ and $(a_s/t)^n$ for $n\geq 1$ can be present in lattice correlators. Such contributions are suppressed at large-Euclidean times, where standard spectroscopic studies are performed, and thus can be safely neglected in such calculations. However, since our goal is an extraction of the logarithmic derivative of the correlation function near $t=0$, the quantification of these possible lattice effects is essential. The exact form of the correlation function at short times is not known, although we found that models of the form,
\begin{align}
\label{eq:model_pole}
G_{\rm mod}(r,t;\vec{\alpha}) = \left(1+\frac{a_t}{t}|C(r;a_t,a_s, N)|\right)\, G_{0,\rm mod}(r,t; \vec{\alpha}) \, ,
\end{align}
were sufficient to describe the data behavior for small $t$. The $G_{0,\rm mod}(r,t; \vec{\alpha})$ function was modeled as a spectral sum of exponentials,
\begin{align}
\label{eq:model_exp}
G_{0,\rm mod}(r,t; \vec{\alpha})=\sum_{n=0}^{n_{\rm exp}} \left|A_n(r;a_t,a_s,N)\right|\,  \text{e}^{-E_n(r; \, a_t, a_s, N) \, t} \, .
\end{align}
To ensure stable extrapolation to $t\to 0$, all fits were performed between $t_{\rm start}$ and $t_{\rm end}$, with \mbox{$t_\text{start}/a_t=1$}. Our model average included contributions from models of the form \cref{eq:model_exp} with varying $n_{\text{exp}}$ and $t_{\rm end}$.  We found that for large anisotropies, the small-time window was modeled well only when at least three exponentials were included in the above sum. However, due to the limited amount of data points available, fits with three exponentials are unstable, and sometimes produce considerable uncertainties when calculating the logarithmic derivative.

We also considered models without explicit pole terms, models with higher order leading poles, and models for which \mbox{$\log \left[ G_{0,\rm mod}(r,t;\vec{\alpha})\right]$} is given by a polynomial. All these models produced seemingly consistent results, but had worse $\chi^2/\text{ndof}$ and were therefore excluded from this study.
    
Examples of the fits to the time-dependence of the correlation function, for $N=24$ lattices,  are shown in~\cref{fig:C(t)_fits}. For a given lattice with $R \times T$ values of the correlation function, for each $R$, we perform a series of model-averaged fits to capture the correlator's $T$-dependence.

\subsubsection{Coulomb potential}
\label{sec:Coulomb-fits}

\begin{figure}[b]
\includegraphics[width=\columnwidth]{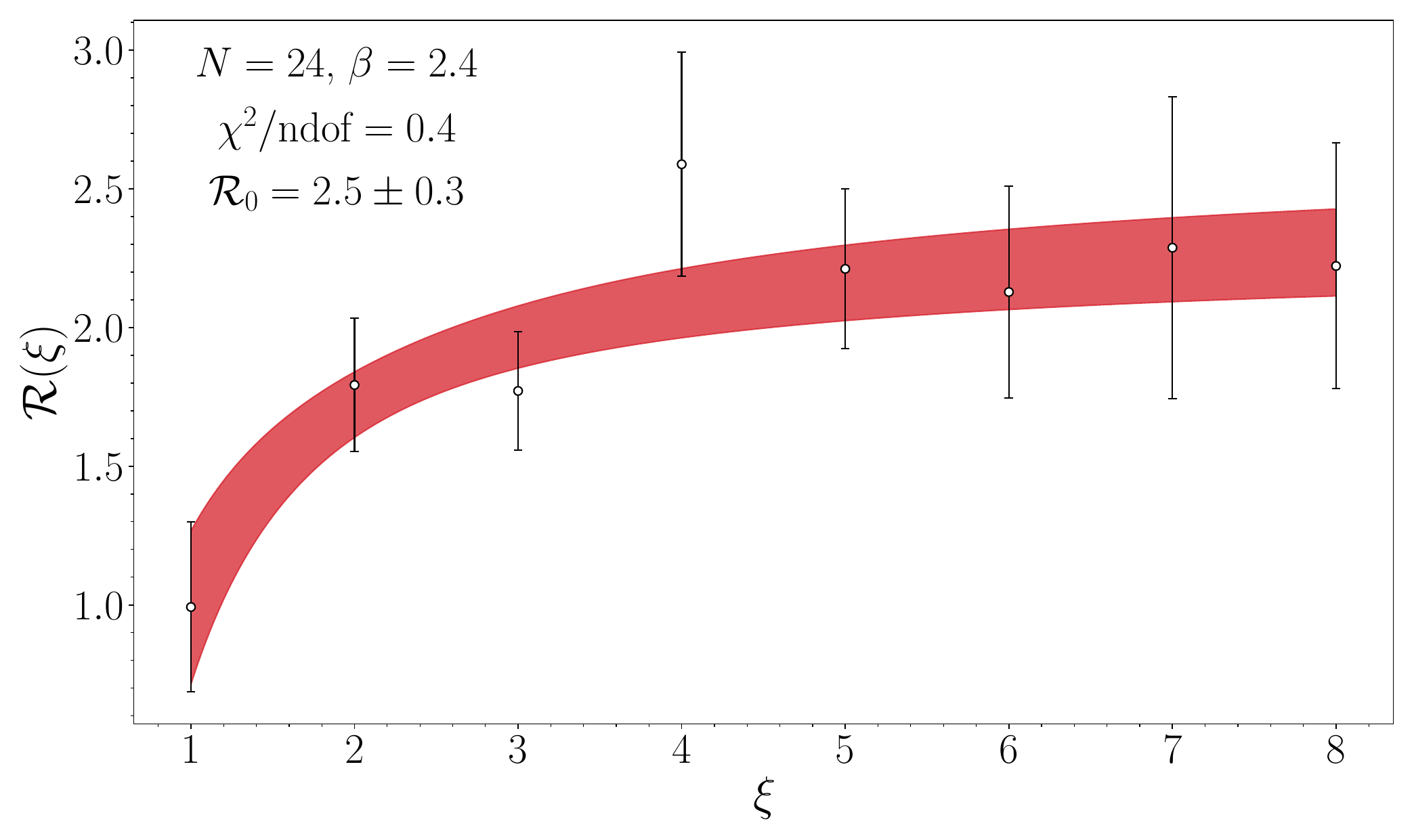} 
\caption{Shown is the extrapolation $\xi\to\infty$ based on the set of eight renormalized anisotropies for $N=24$ and $\beta=2.4$. The filled band region indicates the fitted points for the best model, according to the selection criterion.}
\label{fig:xi_extrapol}
\end{figure} 

\begin{figure*}
\includegraphics[width=\textwidth]{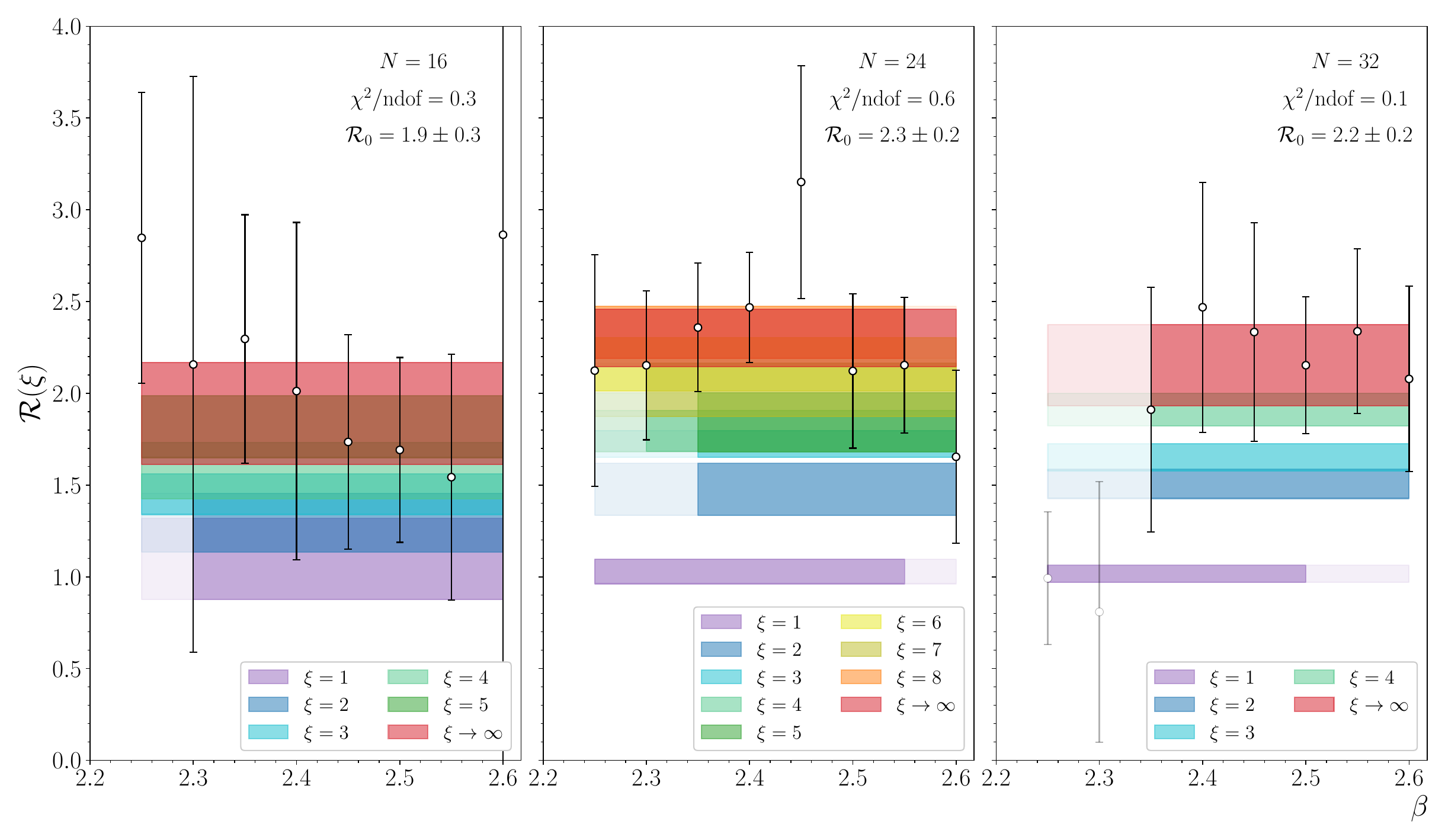}
\caption{Shown are the data points of the extrapolated string tension ratios for every volume under study ($N=16$, $N=24$ and $N=32$), for every $\beta$, and our final fit to a constant in red. The final results and error bands are obtained from model averaging, where the filled region corresponds to the fitted region for the best model selection fit. The remaining color bands represent fits to the corresponding string tension for each anisotropy.}
\label{fig:ratio_fits}
\end{figure*} 

The Coulomb potential was obtained by computing the logarithmic derivative of the correlator at $t=0$,
\begin{align}
V_{\rm lat}(r;a_t,a_s,N)= - \frac{d}{dt} \log \left[ G_{0,\rm mod} \left(r,t; \vec{\alpha} \right) \right] \bigg|_{t=0} \, .
\end{align}
The resulting non-Abelian Coulomb potential as a function of the spatial separation, $r$, for several lattice couplings, is shown in~\cref{fig:V(r)_fits}. To extract the Coulomb string tension we fit the potential to the form of the well-known Cornell form~\citep{Eichten:1978tg},
\begin{align}
V_{\rm lat}(r;a_t,a_s, N) = & -\frac{|A(a_t,a_s,N)|}{r} + B(a_t,a_s,N) \nonumber \\
&  + \sigma_C(a_t,a_s,N) \, r\, ,
\end{align} 
where we interpret $\sigma_C(a_t,a_s,N)$ as the Coulomb string tension at finite lattice spacing in the finite volume. Fits of the $R$-dependence included only data points with $R\leq 3N/8$, to avoid biasing our results with the finite-volume artifacts visible near $R=N/2$.

The use of the Cornell potential in the short-distance region is not particularly well-motivated~\cite {Sommer:1993ce}, due to the known logarithmic corrections to the leading $1/r$ term and the short-range discretization effects. Thus, to estimate the systematic uncertainty associated with this model, we again use model averaging, varying the initial ($R_{\rm start} \leq 1+N/8$) and final ($R_{\rm end}$) ranges of our fits to data so that each fit contains at least $N/4$ points. In doing so, we produce a model averaged value $\sigma_C(a_t,a_s,N)$ for each $\beta$ and $\xi$.

\subsubsection{Hamiltonian limit}
\label{sec:xi_extrapol}

For every ensemble, we calculate the ratio of the Coulomb and physical string tensions,
\begin{align}
\mathcal{R}(\xi) = \frac{\sigma_C(a_t(\xi, \beta), a_s(\xi, \beta), N)}{ \sigma_F(a_t(1, \beta), a_s(1, \beta), N)} \, ,
\end{align}
where we keep the $\beta$ and $N$ dependence implicit. The error propagation in this ratio is treated as uncorrelated.\footnote{Note that the results for $\xi=1$ are correlated, as they come from two different time limits of the same correlator. The ratios for the rest of the anisotropies are indeed uncorrelated. For consistency, we assume all the $\xi-$dependent Coulomb string tensions are independent of the physical ones.} Our $\sigma_F$ is calculated from the large $t$ behavior of two Wilson lines in the Coulomb gauge on isotropic lattices for all considered $\beta$'s. As we demonstrate in~\cref{app:Wilson_limit}, this produces a string tension consistent with the standard computation. We assume that for the ensembles considered in this work, the residual dependence of this ratio on $a_s$ is negligible. We also assume that the limit $t \to \infty$ string tension $\sigma_F$ entering in the denominator does not depend on the anisotropy, though we stress that the value of the Coulomb string tension does strongly depend on this quantity. In the final step of our analysis, for each lattice of size $N$ and coupling $\beta$, we combine all values of the anisotropy $\xi$ to perform an extrapolation $\xi\to\infty$.  To do so, we fit the ratio to a polynomial expansion in $1/\xi$,
\begin{align}
\label{eq:xi_extrapol}
\mathcal{R}(\xi) = \mathcal{R}_0 + \sum^{N_p}_{i=1} \mathcal{R}_i \, \left(\frac{1}{\xi} \right)^i \, .
\end{align}
We once again perform model averaging of our results, fitting both two- and three-term polynomials, where the minimum number of fitted data points is $N_p+2$, and the first anisotropy is always fixed to $\xi=1$. An example for \mbox{$N=24$} and \mbox{$\beta=2.4$} is shown in~\cref{fig:xi_extrapol}. We find a smooth dependence in $\xi$ and conclude that a low parameter fit to data is sufficient to describe our computed ratios.

We repeated the previously described fitting procedure for all lattice ensembles and extrapolated to infinite anisotropy in units of the Wilson string tension, producing the values shown in~\cref{fig:ratio_fits}, where each $N$ is fitted separately. Note that all volumes can be described with a fit to a \mbox{$\beta-$independent} constant, with a reasonable $\chi^2/\text{ndof}$ value.  Furthermore, all three $N$ produce compatible results for the ratio of string tensions. We note that the ratios obtained at low $\beta$'s for $N=32$ are disfavored by our model-averaging procedure, and lie below the final value. The fit that includes all ratios obtained from the $N=32$ ensembles still exhibits a reasonable $\chi^2/\text{ndof}$, although it produces a slightly lower final value for the ratio. In this work, we neglect the finite-volume effects on this ratio. Considering that for each $N$ we use a different number of anisotropies to extrapolate, we choose to take an envelope to determine a conservative estimate for our final result as $\mathcal{R}_0 \equiv \sigma_C/\sigma_F = 2.0 \pm 0.4$. Alternatively, when fitting all data points, for all $N$, we obtain \mbox{$\mathcal{R}_0=2.07\pm0.10$}, with a $\chi^2/\text{ndof}=1.0$, in agreement with the more conservative estimate.

\section{Conclusions}
\label{sec:con}

In this paper, we revisited the lattice definition of the non-Abelian Coulomb potential and argued that previous lattice studies may have overestimated the corresponding Coulomb string tension. We proposed an improved method for determining this quantity and found \mbox{$\sigma_C/\sigma_F=2.0 \pm 0.4$}. To the best of our knowledge, this study is the first to account for short-time lattice artifacts when extracting the non-Abelian Coulomb potential. Consequently, our result is significantly smaller than the previously reported values. This reduction may have considerable phenomenological implications, indicating increased feasibility of constructing meson models based on the Coulomb gauge Hamiltonian formalism. The situation for the SU(3) YM theory needs further investigation, as we expect the currently available values in the literature to be similarly inflated.

\acknowledgments
We thank \mbox{M.~Baker}, \mbox{G.~Burgio}, \mbox{R.~Edwards}, \mbox{J.~Greensite}, and \mbox{S.~Sharpe} for helpful discussions. 
SMD acknowledges the financial support through the U.S. Department of Energy Contract No.~\mbox{DE-SC0011637}. WAS acknowledges the support of the U.S.~Department of Energy ExoHad Topical Collaboration, contract \mbox{DE-SC0023598}. RJP has been supported by the projects \mbox{CEX2019-000918-M} (Unidad de Excelencia ``María de Maeztu''), \mbox{PID2020-118758GB-I00}, financed by \mbox{MICIU/AEI/10.13039/501100011033/} and FEDER, UE, as well as by the EU \mbox{STRONG-2020} project, under the program \mbox{H2020-INFRAIA-2018-1} Grant Agreement \mbox{No.~824093}. CFR acknowledges the support of Spanish Ministerio de Ciencia, Innovación y Universidades \mbox{(MICIU)} through Grant \mbox{No.~BG20/00133}. ES acknowledges support from the U.S. Department of Energy under contract \mbox{DE-SC0019232}. This work was supported by the U.S. Department of Energy contract \mbox{DE-AC05-06OR23177}, under which Jefferson Science Associates, LLC operates Jefferson Lab, by the U.S. Department of Energy Grant No.~\mbox{DE-FG02-87ER40365}, and contributes to the aims of the U.S. Department of Energy \mbox{ExoHad} Topical Collaboration, contract \mbox{DE-SC0023598}.

This research was supported in part by Lilly Endowment, Inc., through its support for the Indiana University Pervasive Technology Institute. Furthermore, this work was supported by the Research Computing clusters at Old Dominion University. The Wahab cluster at Old Dominion University is supported in part by National Science Foundation's grant \mbox{CNS-1828593}. We also acknowledge the computational resources and assistance provided by the Centro de Computación de Alto Rendimiento CCAR-UNED.


\appendix

\section{Enhancement of the Coulomb Potential}
\label{app:enhancement}

Let us consider the correlator $G_{\rm lat}(r, t; a) = G_{\rm lat}(r, t; a, a, \infty )$. Here, we fix \mbox{$a = a_s = a_t$}, and take the limit $N \to \infty$ to simplify the argument. Assuming $G_{\rm lat}(r, t; a)$ is a regular function at $t=0$, one may expand it in $t$ about $t=0$ as follows, 
\begin{align}
G_{\rm lat}(r, t; a) = G(r,0;a) + t \, G'(r, 0; a) + \frac{t^2}{2} \, G''(r,0;a) \, ,
\end{align}
up to terms $\Oc(t^3)$. However, this expansion does not account for the short-distance artifacts from the discretization of the lattice in the temporal direction, as discussed in~\cref{sec:potential}. By dimensional analysis, we expect finite-spacing effects to appear in $G_{\rm lat}(r,t; a)$ of the form,
\begin{align}
\label{eq:discrete-corrections}
\Delta G_{\rm lat}(r,t;a) = \sum_{n=1} \left(\frac{a}{t}\right)^n G_n(r, t) \, ,
\end{align}
since $t$ is the dominant scale in the problem\footnote{For simplicity, we neglect terms proportional to $a/r$ in this reasoning.}. Here, $G_n$'s are coefficients that generally depend on $t$ and are expected to decay exponentially with time, just like the continuum correlator. Without directly computing the correlator from the path integral of Coulomb-gauge fixed YM theory, one cannot conclusively argue that the leading discretization artifact is of the order $a$. Similarly, one cannot easily assert that the infinite sum in \cref{eq:discrete-corrections} is finite or infinite at $t=0$. Nevertheless, whether the leading order correction is regular or divergent at vanishing time extent, the correction $\Delta G$ influences the normalization of $G$ at small $t$ in a discretized theory. Taking these discretization effects into account, we write, 
\begin{align}
\label{eq:correlator-corrected}
G_{\rm lat}(r, t; a) &= G(r,0) + t \, G'(r, 0) \nonumber \\
&+ \frac{t^2}{2} \, G''(r,0) + \Delta G(r,t;a) + \Oc(t^3) \, ,
\end{align}
in which the Taylor-expansion coefficients are now written in terms of continuum quantities and do not depend on $a$. We apply the logarithm to the above expression, set $t=a$, and divide by $a$ to recover~\cref{eq:latt-coul-wrong}. Keeping the lowest terms in $a$ yields,
\begin{align}
\widetilde{V}_{\rm lat}(r; a, a, \infty) = - \frac{1}{a} \biggl[& \log \big[ G(r, 0) + \Delta G(r,a;a) \big] \phantom{\biggl]} \nonumber \\
& \! + a \, \frac{G'(r,0)}{G(r,0)} + \Oc\left(a^2\right) \biggr] \, .
\end{align}
Notably, the second term in the above expansion, \mbox{$ - G'(r,0)/G(r,0)$}, is the expected continuum result. The first term vanishes if \mbox{$G(r, 0) + \Delta G(r,a;a) \to 1$} rapidly enough as $a \to 0$. However, even if the continuum correlator is normalized exactly so that \mbox{$G(r,0) = 1$}, the lattice correction $\Delta G(r,a;a)$ makes the first term of the above expansion non-zero since it is evaluated at time $t=a$ comparable with the lattice spacing $a$. For small but finite $a$, the logarithmic term may dominate over the second one, inflating the measured potential as $a$ decreases, as previously observed in Ref.~\citep{Smith:2023pua}.

Therefore, we conclude that one cannot use the $\widetilde{V}_{\rm lat}$ definition,~\cref{eq:latt-coul-wrong}, to compute the continuum non-Abelian Coulomb potential. Instead, one may measure the expectation value of the lattice correlator and directly fit its time dependence near $t=0$ with a well-chosen continuous model, as described in the main text.

\section{Lattice anisotropy}
\label{app:anisotropy}

\begin{figure}
\includegraphics[width=0.48\textwidth]{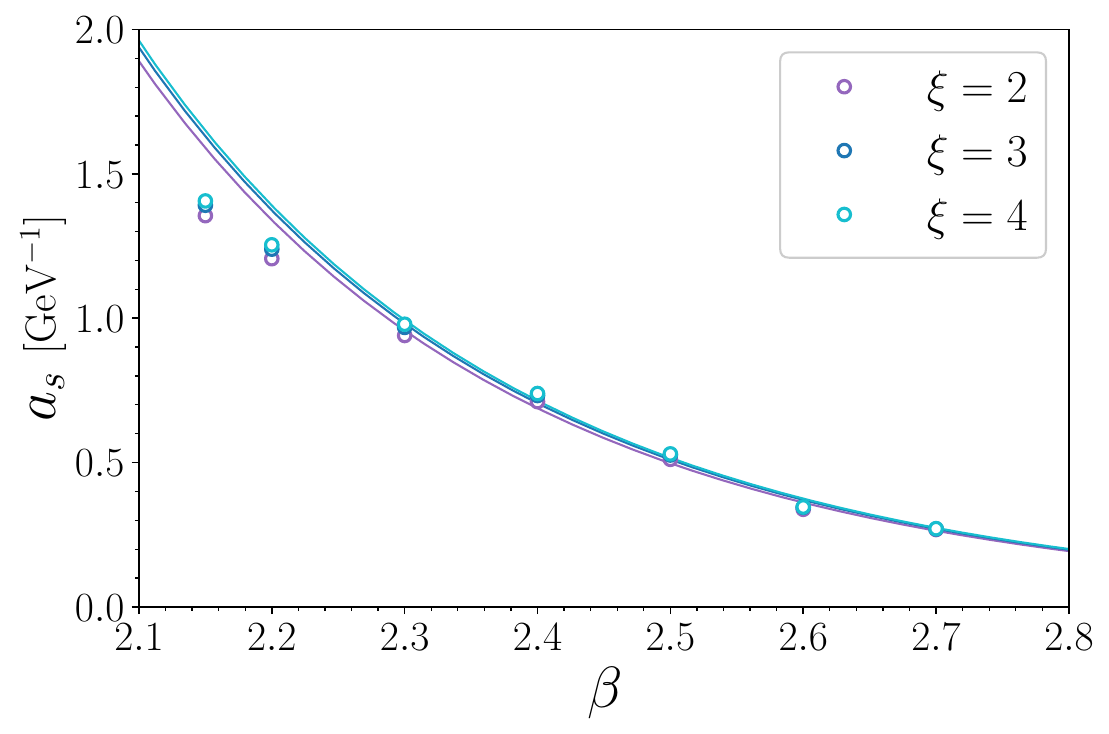}
\caption{Spatial lattice spacing dependence on $\beta$ and the renormalized asymmetry $\xi$. Data from Ref.~\citep{Burgio:2012bk}.}
\label{fig:as}
\end{figure}

To determine dependence of the bare anisotropy $\xi_0$ and spacings $a_t, a_s$ on the renormalized anisotropy $\xi$ and on the lattice coupling $\beta$, one typically computes spatial-spatial, $W_{ss}$, and spatial-temporal Wilson loops, $W_{st}$, tuning $\xi_0$ so that for large enclosed areas, the ratio of the two yields the desired renormalized anisotropy~\citep{Klassen:1998ua, Ishiguro:2001}. This procedure, in principle, requires many additional lattice measurements for each $\beta$ and $\xi$ at which we wish to determine the Coulomb string tension. Hence, to save computer time, we employ a parameterization of $\xi_0(\xi,\beta)$ and $a_s(\xi, \beta)$ based on data provided in Refs.~\citep{GarciaPerez:1996ft, Burgio:2012bk}. We extrapolate the obtained parametrization to values of $\beta$ and $\xi$ not considered in these studies. The bare and the renormalized asymmetries can be related via~\citep{Klassen:1998ua},
\begin{align}
\label{eq:xiR}
\frac{\xi}{\xi_0} \equiv   \eta(\xi, \beta) =  1 + \frac{4}{6} \left(  \frac{1+ a_1 / \beta}{1 + a_2/ \beta} \right) \frac{\eta_1(\xi)}{\beta} \, ,
\end{align}
where $a_1$ and $a_2$ are free parameters and $\eta_1(\xi)$ is the one-loop anisotropy which was computed in~\citep{GarciaPerez:1996ft} and can be parameterized as,
\begin{align}
\label{eq:eta1}
\eta_1(\xi) = c_0 + \frac{c_1}{\xi} + \frac{c_2}{\xi^3} \, ,
\end{align}
where \mbox{$c_0 = 0.3398$}, \mbox{$c_1 = -0.3068$}, \mbox{$c_2 =-0.049$}. Once $\eta_1(\xi)$ is fixed we determine $a_1$ and $a_2$ in~\cref{eq:xiR} fitting the data from Ref.~\citep{Burgio:2012bk} for $\beta>2$, obtaining \mbox{$a_1 = - 1.0238$} and \mbox{$a_2 = -1.8921$}.

The physical spacing, $a_s(\beta,\xi)$  (in units of [GeV$^{-1}$]) is obtained by fitting the renormalization-group inspired parameterization~\citep{Bloch:2004} to the data of Ref.~\citep{Burgio:2012bk}, 
\begin{align}
\sigma_F \, a_s^2(\xi,\beta) = f_1(\xi)^2 \, \text{e}^ {f_2(\beta)} \, ,
\end{align}
where, following Ref.~\citep{Burgio:2012bk}, we set here \mbox{$\sqrt{\sigma_F}=\sqrt{\sigma} = 440$}~MeV and,
\begin{align}
f_1(\xi) = & \, b_0 + \frac{b_1}{\xi} \, ,\\
f_2(\beta) = & \, \frac{2 \beta_1}{\beta_0^2} \log \left(\frac{4 \pi ^2 \beta}{\beta_0} \right) - \frac{ 4 \pi ^2 \beta }{\beta_0} + \frac{ 4 \pi ^2 }{\beta_0}\frac{b_2}{\beta} + b_3 \, ,
\end{align}
where $\beta_0 = 22/3$, $\beta_1 = 68/3$. Parameter $b_0$ is fixed to \mbox{$b_0=c_0=0.3398$} as in Eq.~\eqref{eq:eta1}, and the parameters $b_1$, $b_2$, and $b_3$ are fitted to the $a_s$ provided in Ref.~\citep{Burgio:2012bk} for \mbox{$\beta>2.2$}, obtaining \mbox{$b_1=-0.0477$}, \mbox{$b_2 = 1.6139$}, and \mbox{$b_3=7.0576$}. The result of the fit is shown in~\cref{fig:as}. We note that this parameterization is appropriate for \mbox{$\beta > 2.2$}. 

We verify the correctness of our parametrization for several arbitrarily chosen lattice ensembles. Following Refs.~\citep{Klassen:1998ua, Ishiguro:2001}, we performed explicit calculations of the physical lattice spacings through calculation of $W_{ss}$ and $W_{st}$, finding good agreement of the extracted renormalized anisotropy and spacing with our parameterizations. Finally, we note that our parameterization of $\xi$ and $a_s$ are not appropriate for the isotropic lattices, $\xi=1$, and for this case, we use a parameterization of Ref.~\citep{Bloch:2004}.

\section{Model averaging implementation}
\label{app:MA}

The approach summarized in this section follows from Refs.~\citep{Jay:2020jkz, Neil:2022joj, Neil:2023pgt}. Given a collection of data samples and models\footnote{For simplicity, we describe different functional fitting forms, but also the same functions applied to different data ranges, as different ``models''.}, model averaging enables us to predict a more reliable estimation of our observables of interest. Observables are thus given by\footnote{$a_0$ is to be considered a parameter or observable of interest.}
\begin{align}
\left\langle a_0\right\rangle_{MA}=\sum_M\left\langle a_0\right\rangle_M \operatorname{pr}(M | D),
\label{eq:ma_master}
\end{align}
where the subscript $M$ describes the jackknife average for a given model
\begin{align}
\left\langle a\right\rangle_M=\frac{1}{N}\sum^N_{n=1} a_n,
\label{eq:jackk_avg}
\end{align}
where $N$ is the number of jackknife samples (not to be confused in this appendix with the lattice size), and $a_n$ is the value obtained for each sample. The $M$ model's probability is given by $\operatorname{pr}(M | D)$ when fitted to data $D$. In this work, we make a minimalistic use of priors in the fits, and neglect their contributions thereafter, so that the model weight is given by
\begin{align}
\operatorname{pr}(M | D) & =\frac{\text{e}^{-{\rm AIC}(M|D)/2}}{\sum_{M^\prime} \text{e}^{-AIC(M^\prime|D)/2}},\\
\text{AIC}(M|D)  &=\chi^2(M|D)+2 k +2 n_D,
\end{align}
where AIC stands for a modified version of the Akaike information criterion~\citep{Jay:2020jkz} that also considers models for which data points have been dropped from the full data sample. Here, $\chi^2(M|D)$ is the total $\chi^2$ obtained from the fit, $k$ is the number of free parameters from model $M$, and $n_D$ is the number of dropped data points from $D$. The probabilities have been normalized so that $\sum_M \operatorname{pr}(M | D)=1$. The total $\chi^2$ can be obtained from the fit to the data, obtained over all samples, or approximated from the different jackknife fits $\{\chi^2_n\}$, given by
\begin{align}
\chi^2\simeq \frac{1}{N}\sum^N_{n=1} \chi^2_n-\frac{(N_D-k)}{N-1},
\label{eq:jackk_chi2}
\end{align}
where $N_D$ is the number of fitted data points. With the information above, we can estimate the model-averaged covariance matrix as
\begin{widetext}
\begin{align}
\Sigma(a_0,a_1)_{MA}=\langle a_0 a_1 \rangle -\langle a_0\rangle \langle a_1\rangle & =\sum_{M_0 M_1} \operatorname{pr}(M_0 M_1 | D_0 D_1)\langle a_0 a_1\rangle_{M_0 M_1} \nonumber \\
&-\left(\sum_{M_0 M_1} \operatorname{pr}(M_0 M_1 | D_0 D_1)\langle a_0\rangle_{M_0 M_1})\right)\left(\sum_{M_0 M_1} \operatorname{pr}(M_0 M_1 | D_0 D_1)\langle a_1\rangle_{M_0 M_1}\right),
\label{eq:ma_cov}
\end{align}
where $\sum_{M_0 M_1}\operatorname{pr}(M_0 M_1 | D_0 D_1)$ reduces to $\sum_{M_0}\operatorname{pr}(M_0 | D_0)$ if both observables come from the same model and data combination (for example, when evaluating the slope and intercept of the same line's fit). The variance reduces to
\begin{align}
\Sigma_{MA}=\langle a^2 \rangle -\langle a\rangle^2 & =\sum_{M} \operatorname{pr}(M | D)\langle a^2\rangle_{M}  -\left(\sum_{M} \operatorname{pr}(M | D)\langle a\rangle_M\right)^2 \nonumber \\
& = \sum_M \sigma_{a, M}^2 \operatorname{pr}\left(M | D\right)+\sum_M\left\langle a\right\rangle_M^2 \operatorname{pr}\left(M | D\right)-\left(\sum_M\left\langle a\right\rangle_M\operatorname{pr}\left(M | D\right)\right)^2 ,
\label{eq:ma_variance}
\end{align}
where the first term in the last line accounts for model-averaged statistical errors, and the remaining two describe the systematic spread between models, averaged.

In this work, we assume that the probabilities for different models factorize $\operatorname{pr}(M_0 M_1 | D_0 D_1)=\operatorname{pr}(M_0 | D_0)\operatorname{pr}(M_1 | D_1)$, unless $M_0=M_1$,\, $D_0=D_1$. Note that, making use of~\cref{eq:jackk_avg}, we can interchange the order of the model averaging and jackknife procedures
\begin{align}
\left\langle a\right\rangle_{MA}=\sum_M\left\langle a\right\rangle_M \operatorname{pr}(M | D)=\sum_M\sum^N_n \frac{1}{N} a_n \operatorname{pr}(M | D)=\frac{1}{N}\sum^N_{n=1} \sum_M a_n \operatorname{pr}(M | D)=\frac{1}{N}\sum^N_{n=1} a(M|D)_n.
\end{align}

\begin{figure*}
\includegraphics[width=\textwidth]{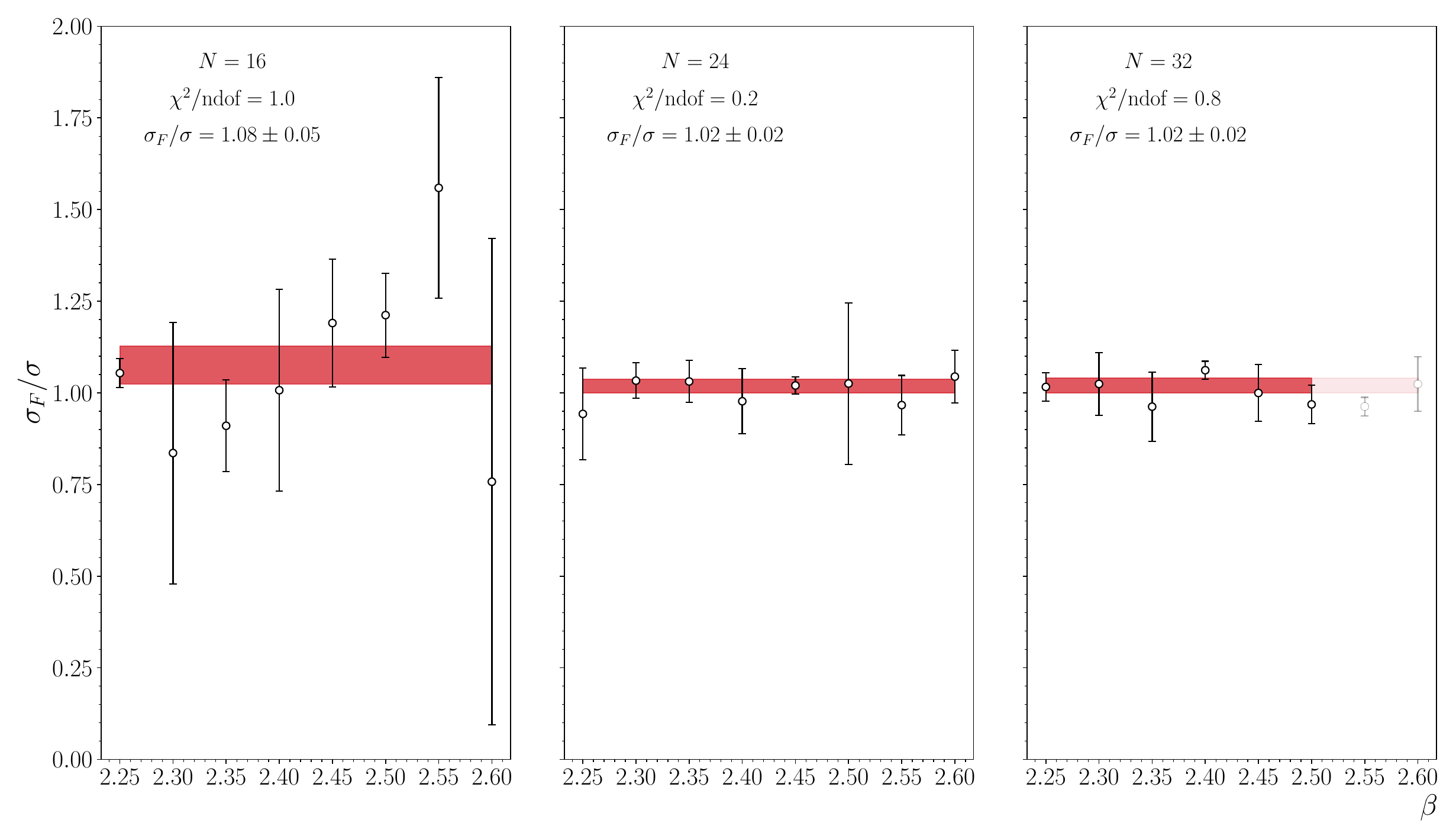} 
\caption{Shown are our values of $\sigma_F$ calculated from long Wilson lines in the Coulomb gauge, divided by the customary string tension value $\sqrt{\sigma}=440$ MeV, for every ensemble in our study. The final results and error bands are obtained from model averaging, where the filled region corresponds to the fitted region for the best model selection fit.}
\label{fig:WL_ratio_fits}
\end{figure*} 

Coincidentally, our approximated non-diagonal covariance elements for different models behave in the same way (for $a_0\neq a_1$,\,$M_0\neq M_1$),
\begin{align}
\Sigma(a_0,a_1)_{MA}&=\sum_{M_0 M_1} \operatorname{pr}(M_0 M_1 | D_0 D_1)\sum^N_{n=1} \frac{1}{N} (a_0 a_1)_n-\left(\sum_{M_0} \operatorname{pr}(M_0 | D_0) \sum^N_{n=1} \frac{1}{N} (a_0)_n \right)\left(\sum_{M_1} \operatorname{pr}(M_1 | D_1)\sum^N_{n=1} \frac{1}{N} (a_1)_n\right) \nonumber \\
&=\frac{1}{N}\sum^N_{n=1} a_0(M_0|D_0)_n \,a_1(M_1|D_1)_n -\left(\frac{1}{N}\sum^N_{n=1} a_0(M_0|D_0)_n\right)\left(\frac{1}{N}\sum^N_{n=1} a_1(M_1|D_1)_n\right).
\end{align}
\end{widetext}

Consequently, one can compute a model-averaged jackknife set of samples and obtain mean values and non-diagonal covariances from the usual definitions for the jackknife. However, this is not true when computing the variances as the order of operation is relevant; for these, one should directly use~\cref{eq:ma_variance}.

Finally, to reduce the contribution of small-weight models that deviate substantially from the mean, models are ordered from highest to lowest probability. Then, we select those models for which the cumulative weight is less than or equal to $90\%$ of the total.

\section{String tension in the $t\to\infty$ limit}
\label{app:Wilson_limit}

In this Appendix, we summarize our calculation for the string tension in the Wilsonian limit. The fitting procedure for the limit $t\to \infty$ proceeds similarly to our fits of the $t\to 0$ limit. However, to determine the ground-state string tension we only use our isotropic lattices, as the quantity $\sigma_F$ is expected to have negligible $\xi$ dependence. Our main fitting model for the correlation functions is  a sum of exponentials with no pole terms,
\begin{align}
G_{\rm mod}(r,t;\vec{\alpha})=G_{0,{\rm mod}}(r,t;\vec{\alpha}) \, ,
\end{align}
where \mbox{$G_{0,{\rm mod}}(r,t;\vec{\alpha})$} is defined in~\cref{eq:model_exp}. We again perform fits and model average, where now both $t_{\rm start}$ and $t_{\rm end}$ are allowed to vary. Following the same steps described in the main text, we compute the potential in the limit $t\to\infty$,
\begin{align}
\label{eq:V_Coulomb}
V^W_{\rm lat}(r;a_t,a_s,N)=-\frac{d}{dt} \, \log(G_{0,{\rm mod}} \left(r,t;\vec{\alpha})\right) \bigg|_{t\to\infty} \, .
\end{align}
Finally, we obtain the Wilson string tension for each isotropic lattice by fitting this potential to the usual Cornell model described above. We stress that no extrapolation to $\xi\to\infty$ is performed here, and the value obtained for $\xi=1$ is used to obtain $\sigma_F$ for all inverse couplings. Our envelope estimate for the Wilson string tension is  $\sqrt{\sigma_F}= 454\pm 14$ MeV, while a combined fit to all data points, for all $N$, produces $\sqrt{\sigma_F}= 444\pm 2$ MeV. Our measurements of $\sigma_F$ are close to the standard value in the literature, $\sqrt{\sigma}=440$ MeV, and compatible with the recent result from Ref.~\citep{Bulava:2024jpj}.

\bibliographystyle{apsrev4-1}
\bibliography{coulomb}

\end{document}